\documentclass[12pt]{article}
\newcommand{\be}{\begin{equation}}
\newcommand{\ee}{\end{equation}}
\newcommand{\bea}{\begin{eqnarray}}
\newcommand{\eea}{\end{eqnarray}}
\newcommand{\sptwo}{1.4}
\newcommand{\doublespace}{\edef\baselinestretch{\sptwo}\Large\normalsize}
\newcommand{\newsection}[1]{\section{#1}\setcounter{equation}{0}}

\newcounter{newapp}
\begin{document}
\vspace*{0.2in}
\begin{center}
{\large\bf Gravitating p-Branes}
\end{center}
\vspace{0.2in}
\begin{center}
{T.E. Clark}\footnote{e-mail address: clark@physics.purdue.edu}$~^a~,~${S.T. Love}\footnote{e-mail address: loves@physics.purdue.edu}$~^{a}~,~${Muneto Nitta}\footnote{e-mail address: nitta@phys-h.keio.ac.jp}$~^b~,~${T. ter Veldhuis}\footnote{e-mail address: terveldhuis@macalester.edu}$~^{a,c}~,~${C. Xiong}\footnote{e-mail address: xiong@purdue.edu}$~^a$\\
\end{center}
\begin{center}
{{\bf a.}~\it Department of Physics,\\
 Purdue University,\\
 West Lafayette, IN 47907-2036, U.S.A.}\\
\end{center}
\begin{center}
{{\bf b.}~\it Department of Physics,\\
 Keio University,\\
 Hiyoshi, Yokohoma, Kanagawa, 223-8521, Japan}\\
\end{center}
\begin{center}
{{\bf c.}~\it Department of Physics \& Astronomy,\\
 Macalester College,\\
 Saint Paul, MN 55105-1899, U.S.A.}
\end{center}
~\\
~\\
\begin{center}
{\bf Abstract}
\end{center}
Coset methods are used to construct the action describing the dynamics associated with the spontaneous breaking of the local Poincar\'e symmetries of $D$ dimensional space-time due to the embedding of a $p$-brane with codimension $N=D-p-1$.  The resulting action is an $ISO(1,p+N)$ invariant form of the Einstein-Hilbert action, which, in addition to the gravitational vielbein, also includes $N$ massive gauge fields corresponding to the broken space translation symmetries which together carry the fundamental representation of the unbroken $SO(N)$ gauge symmetry and an $SO(N)$ Yang-Mills field localized on the brane. The long wavelength dynamics of the gravitating $p$-brane is the same as the action of an $SO(N)$ vector massive Proca field and a non-Abelian $SO(N)$ Yang-Mills field all coupled to gravity in $d=(1+p)$ dimensional space-time.  The general results are specialized to determine the effective action for the gravitating vortex solution in the $D=6$ Abelian Higgs-Kibble model.
\pagebreak
\doublespace

\newsection{Introduction}

The formation of a brane in a target space spontaneously breaks the space-time symmetries of the bulk down to those of the world volume and its complement.  For a bosonic $p$-brane with codimension $N=D-p-1$ embedded in a $D$ dimensional Minkowski space, the $ISO(1, p+N)$ isometry group of the target space is spontaneously broken to the $ISO(1,p)\times SO(N)$ symmetry group of the $d = (1+p)$ dimensional brane world volume and its covolume, the $N$ dimensions of space orthogonal to the $p$-brane.  Hence there are $N$ broken translation symmetries and $N(1+p)$ broken Lorentz symmetries.  As is typical of spontaneous space time symmetry breaking \cite{Ivanov:1975zq}, however, not all of the broken space time symmetries give rise to independent Nambu-Goldstone degrees of freedom. In fact, there are only $N$ independent Nambu-Goldstone bosons, denoted $\phi^i (x)$, $i=1,2,\ldots,N$,  which can be identified with the broken translation generators. Here $x^\mu$, $\mu= 0, 1,\ldots, p$, are the space-time coordinates parametrizing the $d=(1+p)$ dimensional world volume.  $\phi^i (x)$ describes the covolume oscillations of the $p$-brane into the bulk Minkowski space.  Its long wavelength dynamics is given by the reparametrization invariant volume of the brane times the constant brane tension $\sigma$ as is encoded in the Nambu-Goto action
\be
\Gamma = -\sigma \int d^d x \det{e}=
-\sigma \int d^d x \det{\sqrt{\delta^{ij}-\partial_\mu \phi^i \partial^\mu \phi^j}}.
\label{nga}
\ee
The induced Nambu-Goto vielbein $e_\mu^{~m}$, with $m=0,1,\ldots, p$, is given in the static gauge by
\be
e_\mu^{~m}(x) = \delta_\mu^{~m} +\partial_\mu \phi^i  G_{ij}\partial^m \phi^j  ,
\ee
where the $SO(N)$ symmetric matrix $G_{ij}\equiv H^{-1/2}_{ik}\left[\sqrt{(1-H)}-1\right]_{kl} H^{-1/2}_{lj}$, with the matrix $H_{ij}\equiv (\partial_\mu \phi_i \partial^\mu \phi_j)$ \cite{global}. 

The $p$-brane action is invariant under a nonlinear realization of the target space global isometry group of transformations $ISO(1, D-1)$.  In order to have invariance under general coordinate transformations, additional gravitational fields must be introduced.  The purpose of this paper is to construct the action of the world volume localized gravitational fields when the $p$-brane is embedded in curved space.  In short, the dynamics of the oscillating brane in curved space is described by a world volume localized massless graviton represented by an independent dynamical vielbein $e_\mu^{~m} (x)$ and by a non-Abelian $SO(N)$ world volume localized Yang-Mills field represented by a dynamical field $B^{ij}_\mu (x)$ corresponding to the unbroken $SO(N)$ local rotations. In addition, there are $N$ world volume localized vector fields carrying the fundamental representation of $SO(N)$ and represented by the dynamical fields $A^i_\mu$.  As a consequence of the Higgs mechanism, these vector fields which are associated with the broken space translation generators are massive.  An action governing the dynamics for all these fields is derived in a model independent manner using coset methods in which the target space local symmetry group $ISO(1, D-1)$ is nonlinearly realized.  In section \ref{section2}, the nonlinear local transformations of the Nambu-Goldstone fields are introduced via the coset method \cite{Coleman:sm,West:2000hr}.  The locally covariant Maurer-Cartan one-form building blocks for the invariant action are obtained along with the introduction of the dynamical vielbein and vector fields.  Derivatives of these Maurer-Cartan world volume vectors that are covariant with respect to local Lorentz, Einstein and local $SO(N)$ transformations are defined using the spin, and related affine, and $SO(N)$ connections.  In section \ref{section3}, these covariant derivatives are used to construct the low energy locally $ISO(1, D-1)$ invariant action.  Exploiting the spontaneously broken local translation and Lorentz transformations, the action is transformed to and analyzed in the unitary gauge.  The physical degrees of freedom so obtained are the dynamical world volume vielbein, the $N$-plet massive Proca fields and the $SO(N)$ Yang-Mills fields.

These general results only depend on the symmetry breaking pattern and are independent of the nature of the short distance physics that gives rise to it, whether that be a higher dimensional field theory, string theory, or something else all-together. 
To the extent that they can be described by a four dimensional effective theory at long distance scales, the framework therefore also applies to  specific field theoretical realizations of brane world scenarios, such as domain walls in $D=5$ \cite{Rubakov:1983bb}, vortices in $D=6$ \cite{Giovannini:2001hh}, monopole-like defects in $D=7$ \cite{Gherghetta:2000jf}, and instanton-like defects in $D=8$ \cite{Randjbar-Daemi:1983qa}. The globally invariant action in each of these cases describes in a systematic expansion in terms of the ratio of momentum over the inverse defect width the dynamics of the Nambu-Goldstone degrees of freedom, which are the localized zero modes associated with the spontaneously broken symmetries. Additional massless degrees of freedom not associated with spontaneously broken symmetries as well as light massive degrees of 
freedom\footnote{Light massive degrees of freedom may include those associated with breathing modes in case the space time symmetries are broken by several parallel branes.}
 must be added to the action as matter fields in an invariant way \cite{Clark:2002bh}.  In realistic brane world scenarios such matter fields must include Standard Model fields. 

The locally invariant action obtained by gauging the globally invariant action  \cite{Clark:2005ec} describes the dynamics of the massless gauge fields associated with unbroken local space-time and internal symmetries, including the graviton, and the massive gauge fields associated with the spontaneously broken local space and internal symmetries which have obtained their mass by absorbing the Nambu-Goldstone degrees of freedom through the Higgs mechanism. These form the minimal set of degrees of freedom required to realize all symmetries. Additional massless and light bosonic and fermionic degrees with a  spectrum that depends on the details of the underlying short distance physics again must be added to the action in a prescribed invariant way. Specifically, in section \ref{section4} the general results are used to construct a locally invariant effective action corresponding to a vortex embedded in a $D=6$ dimensional space-time as realized in an Abelian Higgs-Kibble model including gravity \cite{Gherghetta:2000qi}. In this  case therefore $p=3$ and $N=2$. A slight extension of the framework describing just the gravitational degrees of freedom is required as the unbroken $U(1)$ symmetry group is in fact a combination of the $SO(2)$ covolume rotations and the internal local $U(1)$ transformations of the Abelian Higgs-Kibble model.
\pagebreak

\newsection{The Coset Construction \label{section2}}

The presence of the $p$-brane in $D$ dimensional target space spontaneously breaks its symmetry group from $ISO(1,D-1)$ to $ISO(1,p)\times SO(N)$.  The low energy action governing the dynamics of the Nambu-Goldstone modes associated with the symmetry breakdown can be constructed using coset methods.  This technique begins by introducing the coset element $\Omega \in ISO(1,D-1)/SO(1,p)\times SO(N)$ where $SO(1,p)$ corresponds to the Lorentz structure group of transformations of the world volume and the $SO(N)$ to the group of rotations of the covolume of the $p$-brane.  These rotations are taken to be unbroken here.  On the other hand, these symmetries can also be broken.  For example, in section \ref{section4} the $D=6$ Abelian Higgs-Kibble model is considered in which a vortex forms and breaks a linear combination of the $SO(2)$ rotation symmetry of the covolume and the internal $U(1)$ gauge symmetry.  The orthogonal combination of symmetry generators yields a conserved $U_Q(1)$ charge.  

The $ISO(1,D-1)$ generators can be expressed in terms of representations of the unbroken $SO(1,p)\times SO(N)$ Lorentz and covolume rotation subgroup.  The $ISO(1,D-1)$ Poincar\'e symmetry generators are given by $P^{\cal M}$, with ${\cal M}=0,1,\ldots, (p+N)$, for space-time translations and $M^{\cal MN}$ for Lorentz transformations.  The generating Poincar\'e algebra is
\bea
\left[M^{\cal MN} , M^{\cal RS} \right] &=& -i \left( \eta^{\cal MR} M^{\cal NS} -\eta^{\cal MS} M^{\cal NR} +\eta^{\cal NS} M^{\cal MR} -\eta^{\cal NR} M^{\cal MS} \right) \cr
\left[M^{\cal MN} , P^{\cal L} \right] &=& i \left( P^{\cal M} \eta^{\cal NL} - P^{\cal N} \eta^{\cal ML} \right) \cr
\left[P^{\cal M} , P^{\cal N} \right] &=& 0 .
\label{ISO(1,d-1)algebra}
\eea
The world volume $SO(1, p)$ Lorentz generators are identified as $M^{\mu\nu}$, where $\mu, \nu =0,1,\ldots, p$, while the unbroken $SO(N)$ charges are identified as $T^{ij} = M^{(p+i) (p+j)}$, where $i, j =1,2,\ldots ,N$.  The broken bulk Lorentz generators form a world volume Lorentz and $SO(N)$ vector $K^\mu_i = 2 M^{(p+i)\mu}$.  The world volume space-time translation generator is the Lorentz vector, 
$SO(N)$ scalar charge $P^\mu$ while the broken bulk space translation generators form a $SO(N)$ vector and Lorentz scalar charge $Z_i =-P^{(p+i)}$.  The $ISO(1, p+N)$ Poincar\'e algebra, equation (\ref{ISO(1,d-1)algebra}), can be written in this basis of genererators.
The commutation relations between the unbroken generators of the world volume $ISO(1, p)$ Poincar\'{e} algebra and $SO(N)$ algebra
are given by
\bea
\left[M^{\mu\nu} , M^{\rho\sigma} \right] &=& -i \left( \eta^{\mu\rho} M^{\nu\sigma} -\eta^{\mu\sigma} M^{\nu\rho} +\eta^{\nu\sigma} M^{\mu\rho} -\eta^{\nu\rho} M^{\mu\sigma} \right) \cr
\left[M^{\mu\nu} , P^{\lambda} \right] &=& i \left( P^\mu \eta^{\nu\lambda} - P^\nu \eta^{\mu\lambda} \right) \cr
\left[P^\mu , P^\nu \right] &=& 0 \cr
\left[T^{ij} , T^{kl} \right] &=& i \left( \delta^{ik} T^{jl} -\delta^{il} T^{jk} +\delta^{jl} T^{ik} -\delta^{jk} T^{il} \right) \cr
\left[P^\mu , T^{ij} \right] &=& 0 = \left[M^{\mu\nu} , T^{ij} \right] , \label{Algebra1}
\eea
while the broken generators lie in representations of the unbroken symmetries or rotate the unbroken into broken charges so that
\bea
\begin{array}{ll}
\left[M^{\mu\nu} , Z_{i} \right] =0 &
\left[M^{\mu\nu} , K^{\lambda}_i \right] = +i \left( K^\mu_i \eta^{\nu\lambda} - K^\nu_i \eta^{\mu\lambda} \right)\\
\left[T^{ij} , Z_{k} \right] = -i \left( Z_i \delta^{jk} - Z_j \delta^{ik} \right) &
 \left[T^{ij} , K^{\mu{k}} \right] = -i \left( K^{\mu i} \delta^{jk} - K^{\mu j} \delta^{ik} \right) \\
\left[P^\mu , Z_{i} \right] = 0 & 
\left[K^\mu_i , P^\nu  \right] = -2i \eta^{\mu\nu} Z_i .\\
\end{array} & & \nonumber \\
 & & 
\label{Algebra}
\eea
Finally, the commutation relations between broken generators take the form
\bea
\left[Z_i , Z_j \right]  & =  & 0 \nonumber \\
\left[K^\mu_i , K^\nu_j \right] & = & +4i \left( \delta_{ij} M^{\mu\nu}-\eta^{\mu\nu} T^{ij} \right) \nonumber \\
\left[K^\mu_i , Z_j \right] & = & -2i \delta_{ij}P^\mu. \label{Algebra3}
\eea

With these definitions the coset element is given by 
\be
\Omega (x) = e^{ix^\mu  P_\mu} e^{i\phi^i(x)Z_i} e^{iv_i^\mu (x) K^i_\mu},
\label{coset}
\ee
where the world volume coordinates, $x^\mu$, act as parameters for translations in the world volume and are part of the coset.  The fields are also defined as functions of $x^\mu$.  The Nambu-Goldstone fields $\phi^i (x)$ along with $v_i^\mu (x)$ act as the remaining coordinates needed to parameterize the coset manifold $ISO(1,D-1)/SO(1,p)\times SO(N)$. 

Left multiplication of the coset elements $\Omega$ by an $ISO(1,D-1)$ group element $g$ which is specified by local infinitesimal parameters $\epsilon^\mu (x), z^i(x), b_i^\mu(x) , \lambda^{\mu\nu}(x), \theta^{ij}(x)$ so that 
\be
g(x) = e^{i\epsilon^\mu(x) P_\mu} e^{iz(x)Z} e^{ib_i^\mu(x) K^i_\mu} e^{\frac{i}{2} \lambda^{\mu\nu}(x) M_{\mu\nu}}e^{\frac{i}{2} \theta^{ij}(x) T_{ij}},
\ee
results in transformations of the space-time coordinates and the Nambu-Goldstone fields according to the general form \cite{Coleman:sm}
\be
g(x)\Omega(x) = \Omega^\prime(x^\prime) h(x) .
\label{leftmult}
\ee
The transformed coset element, $\Omega^\prime(x')$,  is a function of the transformed world volume coordinates and the total variations of the fields so that
\be
\Omega^\prime (x^\prime) = e^{ix^{\prime\mu}  P_\mu} e^{i\phi^{\prime i}(x^\prime)Z_i} e^{iv^{\prime\mu}_i (x^\prime) K^i_\mu},
\ee 
while $h(x)$ is a field dependent element of the stability group $SO(1,p)\times SO(N)$:
\be
h= e^{\frac{i}{2} \alpha_{\mu\nu}(x) M^{\mu\nu}}e^{\frac{i}{2} \beta_{ij}(x) T^{ij}} .
\label{hele}
\ee

Exploiting the algebra of the $ISO(1,D-1)$ charges displayed in equation (\ref{Algebra1})-(\ref{Algebra3}), along with extensive use of the Baker-Campbell-Hausdorff formulae, the local $ISO(1, D-1)$ transformations are obtained as
\bea
x^{\prime\mu}  &=&  [\eta^{\mu\nu} -\lambda^{\mu\nu}(x)]x_\nu +\epsilon^\mu (x) +2b^\mu_i (x) \phi^i (x)\cr
 & & \cr
\phi^{\prime}_i (x^\prime) &=& [\delta_{ij} +\theta_{ij}  (x)]\phi^j (x) + z_i(x) + 2b^\mu_i (x) x_\mu \cr
 & & \cr
v^{\prime \mu}_i (x^\prime) &=& [\eta^{\mu\nu} - \lambda^{\mu\nu} (x)][\delta_{ij} + \theta_{ij} (x)]v_{\nu j} (x) -b_j^\nu (x) M_{\nu k}^{j \rho}(x) \left[\coth{\sqrt{M (x)}}\right]_{\rho i}^{k \mu} \cr
 & & \cr
\alpha^{\mu\nu}(x) &=& \lambda^{\mu\nu}(x) -4b_j^\lambda (x) [M^{-1/2}(x)]^{j \rho}_{\lambda k}  \cr
 & & \left\{ \left[\tanh{\sqrt{M(x)/4}}\right]^{k \mu}_{\rho i} v^{i\nu} (x) -\left[\tanh{\sqrt{M(x)/4}}\right]^{k \nu}_{\rho i} v^{i\mu} (x)\right\} \cr
\beta_{ij}(x) &=& \theta^{ij}(x) -4b_l^\lambda (x) [M^{-1/2}(x)]^{l \rho}_{\lambda k}  \cr
 & & \left\{ \left[\tanh{\sqrt{M(x)/4}}\right]^{k \mu}_{\rho j} v_{\mu i} (x) -\left[\tanh{\sqrt{M(x)/4}}\right]^{k \mu}_{\rho i} v_{\mu j} (x)\right\} ,\cr
 & &  
\label{variations}
\eea
where the matrix $M$ is defined as
\be
M_{ij}^{\mu\nu} =  4\left[ \eta^{\mu\nu} U_{ij} - 2 v^\mu_j v^\nu_i + W^{\mu\nu}\delta_{ij}\right] ,
\ee
with 
\bea
U_{ij} &=& v^\mu_i \eta_{\mu\nu} v^\nu_j \cr
W^{\mu\nu} &=& v^\mu_i \delta^{ij} v^\nu_j
\label{UW}
\eea
and $\eta_{\mu\nu}$ is the metric tensor for $d$ dimensional Minkowski space having signature $(+1,-1,\ldots,-1)$.  In the above, the space-time indices are raised, lowered and contracted using $\eta_{\mu\nu}$ while the Kronecker $\delta_{ij}$ is used for the $SO(N)$ indices.  Both Nambu-Goldstone fields $\phi^i$ and $v^\mu_i$ transform inhomogeneously under the broken local translations $Z^i$ and broken local Lorentz transformations $K^\mu_i$.  Thus these broken transformations can be used to transform to the unitary gauge in which both $\phi^i$ and $v^\mu_i$ vanish.  This will be done in section 3 in order to exhibit the physical degrees of freedom in a more transparent fashion.

The nonlinearly realized $ISO(1,D-1)$ transformations induce a coordinate and field dependent general coordinate transformation of the world volume space-time coordinates.  From the $x^\mu$ coordinate transformation given above, the general coordinate Einstein transformation for the world volume space-time coordinate differentials is given by
\be
dx^{\prime \mu} = dx^\nu {G}_\nu^{~\mu} (x),
\label{dxprime}
\ee
with ${G}_\nu^{~\mu}(x) = \partial x^{\prime \mu}/\partial x^\nu$.  The $ISO(1,D-1)$ invariant interval can be formed using the metric tensor ${g}_{\mu\nu}(x)$ so that $ds^2 = dx^\mu {g}_{\mu\nu}(x) dx^\nu = ds^{\prime 2} = dx^{\prime \mu} {g}^\prime_{\mu\nu}(x^\prime) dx^{\prime \nu}$ where the metric tensor transforms as 
\be
{g}^\prime_{\mu\nu} (x^\prime) = {G}_\mu^{-1\rho}(x) {g}_{\rho\sigma}(x) {G}_\nu^{-1\sigma}(x) .
\label{gprime}
\ee

The form of the vielbein (and hence the metric tensor) as well as the locally $ISO(1,D-1)$
covariant derivatives of the Nambu-Goldstone boson fields and the spin and $SO(N)$ connections can be extracted from the locally covariant Maurer-Cartan one-form, $\Omega^{-1}D\Omega$, which can be expanded in terms of the generators as 
\bea
\omega & = & \Omega^{-1} D\Omega \equiv \Omega^{-1} (d +i \hat{E})\Omega \cr
 &=& i\left[ \omega^m P_m + \omega_{Zi} Z^i +\omega^m_{Ki} K_{im} +\frac{1}{2}\omega_M^{mn} M_{mn}+\frac{1}{2}\omega_{T}^{ij} T_{ij}\right].
 \label{covMC1forms}
\eea
Here Latin indices $m,n = 0,1,\ldots ,p$, are used to distinguish tangent space local Lorentz transformation properties from world volume Einstein transformation properties which are denoted using Greek indices. In what follows Latin indices are raised and lowered with use of the Minkowski metric tensors, $\eta^{mn}$ and $\eta_{mn}$, while Greek indices are raised and lowered with use of the curved space-time metric tensors, $g^{\mu\nu}$ and $g_{\mu\nu}$.  Since the Nambu-Goldstone fields vanish in the unitary gauge it is useful to exhibit the one-form gravitational fields in terms of their translated form
\be
\hat{E} = e^{+ix^m P_m} E e^{-ix^m P_m} .
\ee
The world volume one-form gravitational fields $E$ have the expansion in terms of the charges as
\be
E= E^m P_m +A_i Z^i + B^m_i K_{im} +\frac{1}{2}\gamma^{mn} M_{mn}+\frac{1}{2}{B}^{ij} T_{ij}  .
\ee
Similarly expanding $\hat{E}$ as 
\be
\hat{E}= \hat{E}^m P_m +\hat{A}_i Z^i + \hat{B}^m_i K_{im} +\frac{1}{2}\hat{\gamma}^{mn} M_{mn}+\frac{1}{2}\hat{B}^{ij} T_{ij} ,
\ee
one finds the various fields are related according to 
\bea
\hat{E} &=& E^m +\gamma^{mn}x_n \cr
\hat{A}_i &=& A_i -2 x_m B^m_i  \cr
\hat{B}^m_i&=& B^m_i \cr
\hat{\gamma}^{mn} &=& \gamma^{mn} \cr
\hat{B}^{ij} &=& B^{ij}. 
\eea

Defining the one-form gravitational fields to transform as a gauge field so that
\be
\hat{E}^\prime (x^\prime) = g(x)\hat{E}(x) g^{-1}(x) -ig(x)dg^{-1}(x),
\ee
the covariant Maurer-Cartan one-form transforms analogously to the way it varied for global transformations:
\be
\omega^\prime(x^\prime) = h(x)\omega (x) h^{-1}(x) +h(x)dh^{-1}(x),
\ee
with $h(x)= e^{\frac{i}{2}[\alpha^{mn} M_{mn} +\beta^{ij} T_{ij}]}$ as given in equation (\ref{hele}) and (\ref{variations}).  Expanding in terms of the $D$ dimensional Poincar\'e charges, the individual one-forms transform according to their local $d$ dimensional Lorentz and $SO(N)$ nature so that
\bea
\omega^{\prime m}(x^\prime) &=&  \omega^n (x)\Lambda^{~m}_{n}(\alpha (x)) \cr
\omega_{Zi}^\prime (x^\prime)&=& R_{ij} (\beta(x))\omega_{Zj} (x)\cr
\omega^{\prime m}_{Ki} (x^\prime)&=&  R_{ij} (\beta (x))\omega^n_{Kj} (x)\Lambda^{~m}_{n}(\alpha (x))\cr
\omega^{\prime mn}_M (x^\prime)&=& \omega^{rs}_M (x)\Lambda^{~m}_{r}(\alpha (x))\Lambda^{~n}_{s}(\alpha (x))-d\alpha^{mn}(x) \cr
\omega^{\prime}_{Tij} (x^\prime)&=& R_{ik} (\beta (x)) R_{jl} (\beta (x)) \omega_{Tkl}(x)-d\beta_{ij}(x) .
\label{oneformvari}
\eea
For infinitesimal transformations, the local Lorentz transformations are $\Lambda^{~m}_{n}(\alpha (x)) = \delta^{~m}_{n} + \alpha^{~m}_{n}(x)$ and the local $SO(N)$ transformations are $R_{ij} (\beta (x))= \delta_{ij} +\beta_{ij} (x)$.   Correspondingly the infinitesimal local $ISO(1,D-1)$ transformations of the gravitational one-forms take the form
\bea
\hat{E}^{\prime m}(x^\prime) &=& \hat{E}^n (x) [\delta_n^{~m} +\lambda_n^{~m} (x)] -d\epsilon^m (x)\cr
& &\qquad +\hat{\gamma}^{m}_{~n}(x)\epsilon^n (x) - 2z_i (x)\hat{B}^m_i (x) +2b^m_i (x)  \hat{A}_i (x)\cr
 & & \cr
\hat{A}^\prime_i (x^\prime) &=& [\delta_{ij} +\theta_{ij} (x)] \hat{A}_j (x) -dz_i (x) \cr
 & &\qquad -\hat{B}_{ij} (x) z_j (x) -2\epsilon_m \hat{B}^m_i (x)  + 2b_{im}(x) \hat{E}^m(x) \cr
 & & \cr
\hat{B}^{\prime m}_i (x^\prime) &=& [\delta_{ij} +\theta_{ij} (x)]\hat{B}^n_j (x)[\delta_n^{~m} +\lambda_n^{~m} (x)]-db^m_i (x) \cr
 & &\qquad + \hat\gamma^{m}_{~n}(x) b^n_i (x)  -\hat{B}_{ij} (x) b^m_j (x)  \cr
 & & \cr
\hat\gamma^{\prime m}_{~n} (x^\prime) &=& \hat\gamma^{r}_{~s} (x)[\delta_r^{~m} +\lambda_r^{~m} (x)][\delta^s_{~n} +\lambda^s_{~n} (x)] -d\lambda^m_{~n} (x) \cr
 & &\qquad -4\left(b^m_{i}(x)\hat{B}_{in} (x)-b_{in}(x) \hat{B}^m_i (x) \right)\cr
 & & \cr
\hat{B}^{\prime}_{ij} (x^\prime) &=& [\delta_{ik} +\theta_{ik} (x)][\delta_{jl} +\theta_{jl} (x)]\hat{B}_{kl} (x) -d\theta_{ij} (x) \cr
 & &\qquad +4\left(b_{im}(x)\hat{B}^m_{j} (x)-b_{jm}(x) \hat{B}^m_i (x) \right).
\eea

Using the Feynman formula for the variation of an exponential operator in conjunction with the Baker-Campell-Hausdorff formulae, the individual world volume one-forms appearing in the above decomposition of the covariant Maurer-Cartan one-form, equation (\ref{covMC1forms}), are secured as 
\bea
\omega^m &=& dx^\mu e_\mu^{~m}\cr
  &=& dx^\mu{\cal E}_\mu^{~n} N_n^{~m} \cr
\omega_{Zi} &=& dx^\mu \omega_{Zi\mu} \cr
 &=& dx^\mu \left[\cosh{\sqrt{4U}}\right]_{ij} {\cal E}_\mu^{~m} \left\{ -\left[U^{-1/2}\tanh{\sqrt{4U}}\right]_{jk}v_{km} + {\cal E}_m^{-1\nu} \left( \partial_\nu\phi_j + A_{j\nu} +B_{jk}\phi_k  \right) \right\}\cr
\omega^m_{Ki} &=& dx^\mu \omega_{Ki\mu}^m \cr
 &=&  \left( dv^n_j -\gamma_{nr}v^r_j+B_{jk} v^n_k \right)\left[\sinh{\sqrt{M}}M^{-1/2}\right]^{jm}_{ni} + B^n_j \left[\cosh{\sqrt{M}}\right]^{jm}_{ni}\cr
\omega_M^{mn} &=& dx^\mu \omega_{M\mu}^{mn}\cr
 &=& \gamma^{mn} -4\left[ v^r_i -\gamma^{rs}v_{is} +B_{ik}v^r_k \right] \cr
 & &\qquad  \left\{ \left[\left(\cosh{\sqrt{M}}-1 \right)M^{-1}  \right]^{im}_{rj} v_j^n  -\left[\left(\cosh{\sqrt{M}}-1 \right)M^{-1}  \right]^{im}_{rj} v_j^m   \right\}\cr
 & &-4 B^r_i\left\{\left[M^{-1/2}\sinh{\sqrt{M}} \right]^{im}_{rj} v^n_j  -\left[M^{-1/2}\sinh{\sqrt{M}} \right]^{in}_{rj} v^m_j  \right\}\cr
\omega_{Tij} &=& dx^\mu \omega_{Tij\mu}\cr
 &=& B_{ij} -4\left( dv^m_l -\gamma^{mr}v_{rl}+B_{lk} v^m_k \right)  \cr
 & &\qquad  \left\{ \left[\left(\cosh{\sqrt{M}}-1 \right)M^{-1}  \right]^{ln}_{mj} v_{ni}  -\left[\left(\cosh{\sqrt{M}}-1 \right)M^{-1}  \right]^{ln}_{mi} v_{nj}   \right\}\cr
 & & +4 B_l^m \left\{\left[M^{-1/2}\sinh{\sqrt{M}} \right]^{ln}_{mj} v_{ni}  -\left[M^{-1/2}\sinh{\sqrt{M}} \right]^{ln}_{mi} v_{nj}  \right\}.\cr
 & & 
\label{MCOne-form}
\eea
The covariant coordinate differential $\omega^m$ is related to the world volume coordinate differential $dx^\mu$ by the vielbein $e_\mu^{~m}$, $\omega^m = dx^\mu e_\mu^{~m}$, which in turn can be written in a factorized form as the product $e_\mu^{~m}= {\cal E}_\mu^{~n} N_n^{~m}$ of the dynamic vielbein ${\cal E}_\mu^{~m}$ and the Nambu-Goto vielbein $N_n^{~m}$, where
\bea
{\cal E}_\mu^{~m} &=& \delta_\mu^{~m} + E_\mu^{~m} + 2\phi_i B^{~n}_{i\mu} \cr
N_n^{~m} &=&  \delta_n^{~m} +v_{nj} \left[U^{-1/2}(\cosh{\sqrt{4U}}-1)U^{-1/2}\right]_{ji} v_i^m  \cr
 & & \qquad\qquad - {\cal E}_n^{-1 \nu} (\partial_\nu \phi_j + A_{j\nu} +B_{jk\nu}\phi_k )\left[U^{-1/2}\sinh{\sqrt{4U}}\right]_{ji} v_i^m .
\label{vng}
\eea

The one-forms and their covariant derivatives are the building blocks of the locally $ISO(1, D-1)$ invariant action.  Indeed a $m^{\rm th}$-rank contravariant local Lorentz and $n^{\rm th}$-rank covariant Einstein tensor as well as a $r^{\rm th}$ rank local $SO(N)$ tensor, $T^{m_1\cdots m_m}_{\mu_1\cdots \mu_n~~i_1 \cdots i_r}$ is defined to transform as \cite{Utiyama:1956sy}
\bea
T^{\prime m_1^\prime\cdots m_m^\prime}_{\mu_1^\prime\cdots \mu_n^\prime~~i^\prime_1 \cdots i^\prime_r}(x^\prime) &=& R_{i^\prime_1 i_1}(\beta (x))\cdots R_{i^\prime_r i_r} (\beta (x))G_{\mu_1^\prime}^{-1\mu_1}(x)\cdots G_{\mu_n^\prime}^{-1\mu_n}(x)\times \cr
 & &\times T^{m_1\cdots m_m}_{\mu_1\cdots \mu_n~~i_1 \cdots i_r}(x)\times \cr
 & &\qquad\qquad \times \Lambda^{~m_1^\prime}_{m_1} (\alpha (x))\cdots \Lambda^{~m_m^\prime}_{m_m} (\alpha (x)). 
\eea
For example, the vielbein transforms as $e_\mu^{\prime m} (x^\prime) = G_{\mu}^{-1\nu}(x)
e_\nu^{~n}(x)\Lambda^{~m}_{n} (\alpha (x))$.  Hence, the vielbein and its inverse can be used to convert local Lorentz indices into world volume indices and vice versa.  Since the Minkowski metric, $\eta_{mn}$, is invariant under local Lorentz transformations the metric tensor 
\be
g_{\mu\nu} = e_\mu^{~m} \eta_{mn} e_\nu^{~n} ,
\ee
is a rank 2 Einstein tensor.  It can be used to define covariant Einstein tensors given contravariant ones.  Likewise, the Minkowski metric can be used to define covariant local Lorentz tensors given contravariant ones.  Similarly $SO(N)$ representations have indices that can be contracted to form lower dimensional representations and invariants using the Kronecker delta $\delta_{ij}$.

Since the Jacobian of the $x^\mu \rightarrow x^{\prime \mu}$ transformation is simply
\bea
d^dx^\prime &=& d^d x ~\det{{G}},  
\eea
it follows that $d^dx^\prime ~\det{e^\prime} (x^\prime) = d^dx ~\det{e} (x)$ since $\det{\Lambda} = 1$.
Thus an $ISO(1, D-1)$ invariant action is constructed as
\be
\Gamma = \int d^d x \det{e(x)} {\cal L}(x),
\ee
with the Lagrangian an invariant ${\cal L}^\prime (x^\prime) = {\cal L}(x)$.  The invariants that make up the Lagrangian can be found by contracting the indices of tensors with the appropriate vielbein, its inverse, the Minkowski metric and the $SO(N)$ Kronecker delta.  For example $\omega_{Zi\mu} \delta_{ij} g^{\mu\nu} \omega_{Zj\nu}$ is an invariant term used to construct the Lagrangian, as is $\omega_{Ki\mu}^m \delta_{ij}g^{\mu\nu}\eta_{mn} \omega_{Kj\nu}^n$.

Besides products of the covariant Maurer-Cartan one-forms, their covariant derivatives can also be used to construct invariant terms of the Lagrangian.  The covariant derivative of a general tensor can be defined using the affine and related spin connections and the $SO(N)$ connection.  Consider the covariant derivative of the Lorentz tensor and $SO(N)$ vector $T^{mn}_i$
\be
\nabla_\rho T^{mn}_i = \partial_\rho T^{mn}_i -\omega_{M\rho r}^{m} T^{rn}_i-\omega_{M\rho r}^{n} T^{mr}_i + \omega_{Tij\rho}T^{mn}_j .
\ee
The inhomogeneous transformation (cf. equation (\ref{oneformvari})) of the spin and $SO(N)$ connections, $\omega_{M\mu}^{mn}$ and $\omega_{Tij\mu}$, respectively, are such that the covariant derivative of $T^{mn}_i$ again transforms homogeneously so that
\be
(\nabla_\rho T^{mn}_i)^\prime =R_{ij} G_\rho^{-1\sigma} (\nabla_\sigma T^{rs}_j)\Lambda_r^{~m}\Lambda_s^{~n} .
\ee
Converting the Lorentz index $n$ to a world index $\nu$ using the vielbein, the covariant derivative for mixed tensors is obtained as
\bea
\nabla_\rho T^{m\nu}_i &\equiv & e_n^{-1\nu} \nabla_\rho T^{mn}_i = \partial_\rho T^{m\nu}_i -\omega_{M\rho}^{mr}T_{ri}^{~\nu} + \Gamma_{\sigma\rho}^\nu T^{m\sigma}_i + \omega_{Tij\rho}T^{m\nu}_j ,
\eea
where the spin connection $\omega_{M\rho}^{mn}$ and $\Gamma_{\sigma\rho}^\nu$ are related according to \cite{Utiyama:1956sy}
\be
\Gamma_{\sigma\rho}^\nu = e_n^{-1\nu} \partial_\rho e_\sigma^{~n} -e_n^{-1\nu} \omega_{M\rho}^{nr} e_\sigma^{~s} \eta_{rs}.
\ee
(Note that this relation also follows from the requirement that the covariant derivative of the vielbein vanishes, $\nabla_\rho e_\mu^{~m} =0$.)
Applying the above to the Minkowski metric Lorentz 2-tensor yields the formula relating the affine connection $\Gamma^\rho_{\mu\nu}$ to derivatives of the metric which is given by
\bea
\nabla_\rho \eta^{mn} &=& \partial_\rho \eta^{mn} -\omega_{M\rho r}^{m} \eta^{rn}-\omega_{M\rho r}^{n} \eta^{mr} \nonumber \\
 &=& -\omega_{M\rho}^{mn} -\omega_{M\rho}^{nm}\cr
 &=& 0 \cr
 &=& e_\mu^{~m}e_\nu^{~n} \nabla_\rho g^{\mu\nu} \cr
 &=& e_\mu^{~m}e_\nu^{~n}\left( \partial_\rho g^{\mu\nu} +\Gamma_{\sigma\rho}^\mu g^{\sigma\nu} + \Gamma_{\sigma\rho}^\nu g^{\mu\sigma} \right).
\eea
The solution to this equation yields the affine connection in terms of the derivative of the metric \cite{Utiyama:1956sy} (the space is torsionless, hence the connection is symmetric $\Gamma^\rho_{\mu\nu} = \Gamma^\rho_{\nu\mu}$)
\be
\Gamma^\rho_{\mu\nu} = \frac{1}{2}g^{\rho\sigma}\left[ \partial_\mu g_{\sigma\nu} +\partial_\nu g_{\mu\sigma} - \partial_\sigma g_{\mu\nu}\right].
\ee

Finally a covariant field strength two-form can be constructed out of the inhomogeneously transforming spin connection $\omega_{M\mu}^{mn}$ as
\bea
F^{mn} &=& d\omega_M^{mn} +\eta_{rs} \omega_M^{mr}\wedge \omega_M^{ns} .
\eea
Expanding the forms yields the field strength tensor
\be
F^{mn}_{\mu\nu} = \partial_\mu \omega_{M\nu}^{mn} -\partial_\nu \omega_{M\mu}^{mn}+\eta_{rs} \omega_{M\mu}^{mr} \omega_{M\nu}^{ns} -\eta_{rs} \omega_{M\nu}^{mr} \omega_{M\mu}^{ns} .
\ee
It can be shown that $F^{mn}_{\mu\nu}=e^{-1n\sigma}e_\rho^{~m}R^\rho_{~\sigma\mu\nu}$ where $R^\rho_{~\sigma\mu\nu}$ is the Riemann curvature tensor
\be
R^\rho_{~\sigma\mu\nu} = \partial_\nu \Gamma^\rho_{\sigma\mu} -\partial_\mu \Gamma^\rho_{\sigma\nu} +\Gamma^\lambda_{\sigma\mu}\Gamma^\rho_{\lambda\nu} -\Gamma^\lambda_{\sigma\nu}\Gamma^\rho_{\lambda\mu} .
\ee
The Ricci tensor is given by $R_{\mu\nu} = R^\rho_{\mu\nu\rho}$ and hence the scalar curvature 
\be
R = g^{\mu\nu} R_{\mu\nu} = - e^{-1\mu}_m e_n^{-1\nu} F^{mn}_{\mu\nu} 
\ee
is an invariant. Similarly the $SO(N)$ covariant field strength two-form can be constructed using the inhomogeneously transforming $SO(N)$ connection $\omega_{T\mu}^{ij}$ to be
\be
F^{ij}= d\omega^{ij}_T +\delta_{kl} \omega^{ik}_T \wedge \omega_T^{jl}  .
\ee
Expanding the forms yields the $SO(N)$ Yang-Mills field strength tensor
\be
F^{ij}_{\mu\nu} = \partial_\mu \omega_{T\nu}^{ij} -\partial_\nu \omega_{T\nu}^{ij}+\delta_{kl} \omega_{T\mu}^{ik} \omega_{T\nu}^{jl} -\delta_{kl} \omega_{T\nu}^{ik} \omega_{T\mu}^{jl} .
\ee
Contracting the world volume indices with the space-time metric $g^{\mu\nu}$ and the $SO(N)$ indices with the Kronecker delta $\delta_{ij}$, the Yang-Mills invariant kinetic energy term is obtained as
\bea
F_{\mu\nu}^{ij} F_{ij}^{\mu\nu} & = & F_{\mu\nu}^{ij}g^{\mu\rho}g^{\nu\sigma} \delta_{ik}\delta_{jl} F_{\rho\sigma}^{kl}.
\eea
\pagebreak

\newsection{The Invariant Action \label{section3}}

The covariant derivatives of the Maurer-Cartan one-forms provide additional building blocks out of which the invariant action is to be constructed.  For example the covariant derivatives of $\omega_{Zi\nu}$ and $\omega_{Ki\nu}^n$ give the mixed tensors
\bea
\nabla_\mu \omega_{Z\nu}^i &=& \partial_\mu \omega_{Z\nu}^i - \Gamma_{\mu\nu}^\rho \omega_{Z\rho}^i +\omega_{T\mu}^{ij}\omega_{Z\nu}^j \cr
\nabla_\mu \omega_{Ki\nu}^n &=& \partial_\mu \omega_{Ki\nu}^n -\Gamma_{\mu\nu}^\rho \omega_{Ki\rho}^n -\omega_{M\mu}^{nr} \omega_{Ki\nu}^s \eta_{rs}+\omega_{T\mu}^{ij}\omega_{Kj\nu} . \label{derivatives}
\eea
It follows that the invariant action describing an oscillating $p$-brane embedded in curved $D$ dimensional space-time has the general low energy form
\bea
\Gamma &=& \int d^d x \det{e} \left\{ \Lambda + \frac{1}{2\kappa^2} R -\frac{Z_N}{8}F_{\mu\nu}^{ij} F^{\mu\nu}_{ij}\right.\cr
 & &\left.+\frac{1}{2} \omega_{Z\mu}^i \left[(M^2 + \xi R)g^{\mu\nu}\delta_{ij} + \zeta R^{\mu\nu}\delta{ij} + \chi F_{ij}^{\mu\nu}\right]  \omega_{Z\nu}^j \right.\cr
 & &\left.  \right. \cr
 & &\left.- \frac{Z_1}{4}\left[\nabla_\mu \omega_{Z\nu}^i - \nabla_\nu \omega_{Z\mu}^i\right]\left[\nabla^\mu \omega_{Zi}^\nu - \nabla^\nu \omega_{Zi}^\mu\right]\right\} . \label{action}
\eea
Many invariant terms are possible. The above action includes a reduced set of terms which are used as an effective theory.  The model can be further simplified by setting the parameters $\xi$ , $\zeta$ and $\chi$ to zero. On the other hand, due to the Higgs mechanism, the parameter $M$ cannot be zero and is an independent scale in the theory.  Hence the gauge fields $A_\mu^i$ corresponding to the broken space translation symmetries become massive Proca fields.  Taken together, they lie in the fundamental representation of the unbroken local $SO(N)$ group as do the $Z_i$ charges.  Moreover, exploiting the identity $\partial_\mu (\det{e}~ T^\mu )= \det{e}~ \nabla_\mu T^\mu$ along with the chain rule for covariant differentiation, integration by parts has been used to eliminate redundant terms.  Further terms containing $\omega_{Ki\mu}^m$ have not been included. Such terms do not give rise to new dynamics since  in unitary gauge, in which the Nambu-Goldstone bosons $\phi^i$ and the auxiliary vector fields $v^i_\mu$ are absent, the gravitational field $B^m_{i\mu}$ does not occur in the action Eq.(\ref{action}). Therefore, if terms at most quadratic  in $\omega_{Ki\mu}^m$ are included, the $B^m_{i\mu}$ field acts as a Lagrange multiplier whose elimination generates terms already present in the action.
 
Exploiting the form of the covariant derivative of the $Z^i$ one-form, equation (\ref{derivatives}), the $N$
anti-symmetric field strength tensors, $F_{\mu\nu}^i$, are given by
\bea
F_{\mu\nu}^i&=&\nabla_\mu \omega_{Z\nu}^i -\nabla_\nu \omega_{Z\mu}^i \cr
 &=& [\partial_\mu \delta^{ij} +\omega_{T\mu}^{ij}] \omega_{Z\nu}^j -[\partial_\nu \delta^{ij}+\omega_{T\nu}^{ij}] \omega_{Z\mu}^j  \cr
 &= & D_\mu \omega_{Z\nu}^i - D_\nu \omega_{Z\mu}^i .
\eea
Here $D_\mu^{ij}  \equiv \partial_\mu \delta^{ij} +\omega_{T\mu}^{ij}$ is the $SO(N)$ partially covariant derivative.
As with the connections $A_\mu^i$, the field strength tensors carry the fundamental representation of the unbroken local $SO(N)$.  Thus the action becomes
\bea
\Gamma &=& \int d^d x \det{e} \left\{ \Lambda + \frac{1}{2\kappa^2} R -\frac{Z_N}{8}F_{\mu\nu}^{ij} F^{\mu\nu}_{ij}\right.\cr
 & &\left.+\frac{1}{2} \omega_{Z\mu}^i \left[(M^2 + \xi R)g^{\mu\nu}\delta_{ij} + \zeta R^{\mu\nu}\delta{ij} + \chi F_{ij}^{\mu\nu}\right]  \omega_{Z\nu}^j \right.\cr
 & &\left.  \right. \cr
 & &\left.- \frac{Z_1}{4} F^i_{\mu\nu} g^{\mu\rho} g^{\nu\sigma}\delta_{ij} F^j_{\rho\sigma} \right\} .
\label{effaction}
\eea

According to equation (\ref{variations}), $\phi^i$ and $v^m_i$ transform inhomogeneously under the broken translation and Lorentz transformation local variations.  Unitary gauge is defined by  $\phi^i =0= v^m_i$. In this gauge, the covariant one-forms take the simplified form
\bea
\omega^m &=& dx^\mu e_\mu^{~m} =dx^\mu {\cal E}_\mu^{~m} = dx^\mu (\delta_\mu^{~m} + E_\mu^{~m}) \cr
\omega_Z^i &=& dx^\mu  A_\mu^i \cr
\omega_{Ki}^m &=& dx^\mu B_{i\mu}^{~m}\cr
\omega_M^{mn} &=& dx^\mu \gamma_\mu^{mn} \cr
\omega_T^{ij} &=& dx^\mu B^{ij}_\mu .
\label{MCOne-formUnitary}
\eea
Note that, in this gauge, equation (\ref{vng}) reduces to ${\cal E}_\mu^{~m} =\delta_\mu^{~m} + E_\mu^{~m}$ and $N_n^{~m} = \delta_n^{~m}$.  Consequently the vielbein is $e_\mu^{~m} = {\cal E}_\mu^{~n} N_n^{~m} = \delta_\mu^{~m} + E_\mu^{~m}$ and thus depends only on the gravitational fluctuation field, $E_\mu^{~m}$, about the flat background vielbein $\delta_\mu^{~m}$ and is independent of the vector fields.  As such, the $\det{e} $ gives no contribution to the $U(1)$ vector mass even though it is the source of the Nambu-Goldstone boson kinetic energy term in the model with spontaneously broken global isometry.  Instead, the mass of the vector, $M$, is a new scale which arises from a new independent monomial. 

In the unitary gauge the action, equation (\ref{effaction}), reduces to that of an $SO(N)$ $N$-plet of massive Proca fields, $A_\mu^i$, corresponding to the broken space translation symmetries, and an $SO(N)$ Yang-Mills field, $B^{ij}_\mu$, corresponding to the local $SO(N)$ rotation invariance of the covolume, both coupled to a gravitational field , $e_\mu^{~m}$, with cosmological constant so that
\bea
\Gamma &=& \int d^d x \det{e} \left\{ \Lambda + \frac{1}{2\kappa^2} R -\frac{Z_N}{8}F_{\mu\nu}^{ij} F^{\mu\nu}_{ij}\right.\cr
 & &\left.- \frac{Z_1}{4} F^i_{\mu\nu} F^{\mu\nu}_i 
+\frac{1}{2} A_{\mu}^i \left[(M^2 + \xi R)g^{\mu\nu}\delta_{ij} + \zeta R^{\mu\nu}\delta{ij} + \chi F_{ij}^{\mu\nu}\right] A_{\nu}^j \right\} ,\cr
 & & 
\eea
with the field strength tensor $F_{\mu\nu}^i$ for the $N$-plet Proca field
\bea
F_{\mu\nu}^i &=& [\partial_\mu \delta^{ij} +B_{\mu}^{ij}] A_{\nu}^j -[\partial_\nu \delta^{ij}+B_{\nu}^{ij}] A_{\mu}^j  \cr
 &\equiv & D_\mu A_{\nu}^i - D_\nu A_{\mu}^i ,
\eea
and the $SO(N)$ Yang-Mills field strength
\bea
F_{\mu\nu}^{ij} & = & \partial_\mu  B_\nu^{ij} -\partial_\nu B_\mu^{ij} + B_\mu^{ik} B_\nu^{jk} - B_\nu^{ik} B_\mu^{jk}.
\eea
The action describes a $p$-brane with codimension N embedded in a space gravitating about a background Minkowski space.  The presence of a world volume cosmological constant further allows for topology changes for it.  The world volume action of the $p$-brane is equivalent to that of a world volume gravitational field Einstein-Hilbert action, with corresponding cosmological constant, and the action for a massive $SO(N)$ $N$-plet Proca field as well as an $SO(N)$ Yang-Mills gauge field in that gravitating space.  
~\\
\newsection{The D=6 Abelian Higgs Model \label{section4}}

The six dimensional Abelian Higgs-Kibble model exhibits the Poincar\'e symmetries of D=6 space-time and an internal $U_{\cal Q}(1)$ gauge symmetry.  The formation of a codimension  2 vortex in D=6 breaks this $ISO(1,5)\times U_{\cal Q}(1)$ symmetry down to the D=4 world volume Poincar\'e symmetries, $ISO(1,3)$, and an unbroken Abelian symmetry, denoted $U_Q(1)$, the generator of which is a linear combination of the generator of rotations in the covolume and the original $U_{\cal Q}(1)$ gauge symmetry charge.  The method of nonlinear realizations can be used to construct the long wavelength effective action for the Nambu-Goldstone bosons self-coupling and their interactions with matter.  In addition, the low energy gravitating vortex effective action can be constructed.  Using the previous notation for a $p=3$ vortex brane with codimension N=2, the linear combination of the $SO(2)$ covolume rotation charge $T^{ij} = \epsilon^{ij} T$ and the gauge $U_{\cal Q}(1)$ charge ${\cal Q}$ corresponding to the vortex broken symmetry is
\be
{\cal Y}\equiv T \cos\hat{\theta} + {\cal Q} \sin\hat{\theta} ,
\ee
with $\hat{\theta}$ a mixing angle fixed by the definition of the charge for the remaining unbroken $U_Q(1)$ combination of charges
\be
Q\equiv -T\sin\hat{\theta} + {\cal Q} \cos\hat{\theta} .
\ee

The coset element $\Omega \in ISO(1,5)\times U_{\cal Q}(1) / SO(1,3)\times U_Q(1)$ can be parameterized as
\be
\Omega (x) = e^{ix^\mu  P_\mu} e^{i\phi^i(x)Z_i} e^{iv_i^m (x) K^i_m} e^{i\eta (x) {\cal Y}} ,
\label{vortexcoset}
\ee
with $\phi^i$, $v^m_i$, and $\eta$ the associated Nambu-Goldstone fields where $i=1,2$ and $\mu , m =0,1,2,3$.  The $ISO(1,5)\times U_{\cal Q}(1)$ locally covariant Maurer-Cartan one forms are defined using equation (\ref{covMC1forms}) as
\bea
\omega & = & \Omega^{-1} D \Omega = \Omega^{-1} (d + i\hat{E} ) \Omega \cr
 &=& i\left[ \omega^m P_m + \omega_{Zi} Z^i +\omega^m_{Ki} K_{im}+\omega_{{\cal Y}} {\cal Y} +\frac{1}{2}\omega_M^{mn} M_{mn} + \omega_Q Q\right],
\eea
where now 
\be
\hat{E}= \hat{E}^m P_m +\hat{A}_i Z^i + \hat{B}^m_i K_{im} +\frac{1}{2}\hat{\gamma}^{mn} M_{mn} + \hat{B} T+\hat\gamma {\cal Q} .
\ee
The last two terms in this expression can be cast in the form
\bea
\hat{B} T +\hat\gamma {\cal Q} &=& (\hat{B} \cos\hat{\theta} + \hat\gamma \sin\hat{\theta} ){\cal Y} +(-\hat{B} \sin\hat{\theta} +\hat\gamma \cos\hat{\theta} ) Q \cr
 &\equiv & B_{\cal Y} {\cal Y} + a Q .
\eea

The transformations of the world volume coordinates and the fields can be determined as before, except now the group transformation element is represented by
\be
g(x) = e^{i\epsilon^\mu(x) P_\mu} e^{iz(x)Z} e^{ib^\mu_i(x) K^i_\mu} e^{\frac{i}{2} \lambda^{\mu\nu}(x) M_{\mu\nu}}e^{iy(x) {\cal Y}} e^{i\alpha (x) Q} .
\ee
The variations and the new stability group element $h(x)=e^{\frac{i}{2} \alpha^{mn}M_{mn}} e^{i\varphi Q}$ are obtained as
\bea
x^{\prime\mu}  &=&  [\eta^{\mu\nu} -\lambda^{\mu\nu}(x)]x_\nu +\epsilon^\mu (x) +2b^\mu_i (x) \phi^i (x)\cr
 & & \cr
\phi^{\prime}_i (x^\prime) &=& [\delta_{ij} + (y(x)\cos\hat{\theta} -\alpha (x) \sin\hat{\theta} )\epsilon_{ij}]\phi^j (x) + z_i(x) + 2b^\mu_i (x) x_\mu \cr
 & & \cr
v^{\prime \mu}_i (x^\prime) &=& [\eta^{\mu\nu} - \lambda^{\mu\nu} (x)][\delta_{ij} + (y(x)\cos\hat{\theta} -\alpha (x) \sin\hat{\theta} )\epsilon_{ij}]v_{\nu j} (x) \cr
 & & \qquad\qquad\qquad\qquad\qquad -b_j^\nu (x) M_{\nu k}^{j \rho}(x) \left[\coth{\sqrt{M (x)}}\right]_{\rho i}^{k \mu} \cr
 & & \cr
\eta^\prime (x^\prime) &=& \eta (x) + y(x) +\beta(x) \cos\hat{\theta} \cr
 & & \cr
\varphi (x) &=& \alpha (x) -\beta (x) \sin\hat{\theta},\cr
 & &  
\label{variationsvortex}
\eea
where $\alpha^{mn} (x)$, $M_{\nu k}^{j \rho}(x)$ and $\beta (x)= (\beta_{12}(x) -\theta_{12}(x))$ are given in equations (\ref{variations})-(\ref{UW}).

Consequently the covariant Maurer-Cartan one forms transform as in equation (\ref{oneformvari}) except that the rotation matrix $R_{ij}$ is given in terms of the $U_Q(1)$ transformation parameter $\varphi$ and the mixing angle $\hat{\theta}$
\be
R_{ij}= \delta_{ij} -\varphi \sin\hat{\theta} \epsilon_{ij} .
\ee
In addition the ${\cal Y}$ one-form is found to be invariant,
\bea
\omega^\prime_{\cal Y} (x^\prime) &=& \omega_{\cal Y} (x) ,
\eea
while the $Q$ connection transforms as 
\bea
\omega^\prime_Q (x^\prime ) &=& \omega_Q (x) -d\varphi (x) .
\eea
Denoting the one forms listed earlier in equation (\ref{MCOne-form}) with a tilde, the new Maurer-Cartan one forms are given by
\bea
\omega^m &=& \tilde{\omega}^{m} \cr
\omega_{Zi} &=& \left( e^{-\eta(x) \cos\hat{\theta} \epsilon_{..}}  \right)_{ij} \tilde{\omega}_{Zj} \cr
\omega^m_{Ki} &=& \left( e^{-\eta(x) \cos\hat{\theta} \epsilon_{..}}  \right)_{ij} \tilde{\omega}^m_{Kj} \cr
\omega^{mn}_M &=& \tilde{\omega}^{mn}_{M} \cr
\omega_{\cal Y} &=& d\eta(x) +\tilde{\omega}_{T12}\cos\hat{\theta} +\hat{\gamma} \sin\hat{\theta} = d\eta + B_{\cal Y} +\cdots \cr
\omega_Q &=& \hat{\gamma} \cos\hat{\theta} -\tilde{\omega}_{T12}\sin\hat{\theta} = a +\cdots .
\eea

Just as before, the one-forms and their covariant derivatives can be used as the building blocks to construct a locally $ISO(1,5)\times U_{\cal Q}(1)$ invariant action.  The covariant derivatives of the one-forms are defined using the spin and $U_Q(1)$ connections, so that, for example, the covariant derivative of a tensor,
\be
T^\prime_i (x^\prime) = R_{ij} (\varphi) T_j (x) = T_i (x) -\varphi \sin\hat{\theta} \epsilon_{ij} T_j (x) ,
\ee
is defined as
\be
\nabla_\mu T_i = \partial_\mu T_i -\omega_{Q\mu} \sin\hat{\theta} \epsilon_{ij} T_j 
\ee
and transforms homogeneously
\be
(\nabla_\mu T_i)^\prime = G^{-1\nu}_\mu R_{ij} \nabla_\nu T_j .
\ee
Using the one-forms and their covariant derivatives, candidates for the invariant Lagrangian can be constructed.  In particular the field strength tensor for the ${\cal Y}$ one-form is given by
\be
F_{{\cal Y}\mu\nu} \equiv \nabla_\mu \omega_{{\cal Y}\nu} -\nabla_\nu \omega_{{\cal Y}\mu} = \partial_\mu \omega_{{\cal Y}\nu} -\partial_\nu \omega_{{\cal Y}\mu}
\ee
and similarly the $Q$ one-form field strength tensor is
\be
F_{Q\mu\nu} \equiv \nabla_\mu \omega_{Q\nu} -\nabla_\nu \omega_{Q\mu} = \partial_\mu \omega_{Q\nu} -\partial_\nu \omega_{Q\mu} .
\ee
Unlike the $Q$ one-form, an invariant \lq\lq mass" term can be constructed for the ${\cal Y}$ one-form
$\omega_{{\cal Y}\mu} g^{\mu\nu} \omega_{{\cal Y}\nu}$.

The gravitating vortex invariant action thus has the form
\bea
\Gamma &=& \int d^4 x \det{e} \left\{ \Lambda + \frac{1}{2\kappa^2} R -\frac{Z_{\cal Y}}{4}F_{{\cal Y}\mu\nu} F^{\mu\nu}_{{\cal Y}}-\frac{Z_Q}{4}F_{Q\mu\nu} F^{\mu\nu}_{Q} -\frac{Z_{{\cal Y}Q}}{2}F_{{\cal Y}\mu\nu} F^{\mu\nu}_{Q}\right.\cr
 & &\left.  \right. \cr
 & &\left.- \frac{Z_1}{4}\left[\nabla_\mu \omega_{Z\nu}^i - \nabla_\nu \omega_{Z\mu}^i\right]\left[\nabla^\mu \omega_{Zi}^\nu - \nabla^\nu \omega_{Zi}^\mu \right]\right. \cr
 & &\left.  \right. \cr
 & &\left.+\frac{1}{2} \omega_{Z\mu}^i \left[(M^2 + \xi R)g^{\mu\nu}\delta_{ij} + \zeta R^{\mu\nu}\delta{ij} + \chi F_{{\cal Y}}^{\mu\nu}\epsilon_{ij} + \rho F_Q^{\mu\nu}\epsilon_{ij} \right]  \omega_{Z\nu}^j \right.\cr
 & &\left.  \right. \cr
 & &\left.+\frac{1}{2} \omega_{{\cal Y}\mu} \left[(m^2 + \xi_{\cal Y} R)g^{\mu\nu} + \zeta_{\cal Y} R^{\mu\nu} \right]  \omega_{{\cal Y}\nu} \right\} .
\eea
As with the ${\cal Y}$ and $Q$ one-forms, a field strength tensor for the $Z_i$ one-forms can be introduced as
\bea
F_{\mu\nu}^i &=& \nabla_\mu \omega_{Z\nu}^i -\nabla_\nu \omega_{Z\mu}^i \cr
 &=& [\partial_\mu \delta^{ij} -\omega_{Q\mu} \sin\hat{\theta} \epsilon_{ij}]\omega_{Z\nu}^j - [\partial_\nu \delta^{ij} -\omega_{Q\nu} \sin\hat{\theta} \epsilon_{ij}]\omega_{Z\mu}^j \cr
 &\equiv& D_\mu \omega_{Z\nu}^i -D_\nu \omega_{Z\mu}^i  ,
\eea
with the partially covariant derviative defined as $D_\mu^{ij} \equiv \partial_\mu \delta_{ij} -\omega_{Q\mu} \sin\hat{\theta} \epsilon_{ij}$.
Written in terms of these field strengths, the action reads
\bea
\Gamma &=& \int d^4 x \det{e} \left\{ \Lambda + \frac{1}{2\kappa^2} R -\frac{Z_{\cal Y}}{4}F_{{\cal Y}\mu\nu} F^{\mu\nu}_{{\cal Y}}-\frac{Z_Q}{4}F_{Q\mu\nu} F^{\mu\nu}_{Q}\right.\cr
 & &\left.  \right. \cr
 & &\left. -\frac{Z_{{\cal Y}Q}}{2}F_{{\cal Y}\mu\nu} F^{\mu\nu}_{Q} - \frac{Z_1}{4}F^i_{\mu\nu} F^{i\mu\nu}\right. \cr
 & &\left.  \right. \cr
 & &\left.+\frac{1}{2} \omega_{Z\mu}^i \left[(M^2 + \xi R)g^{\mu\nu}\delta_{ij} + \zeta R^{\mu\nu}\delta{ij} + \chi F_{{\cal Y}}^{\mu\nu}\epsilon_{ij} + \rho F_Q^{\mu\nu}\epsilon_{ij} \right]  \omega_{Z\nu}^j \right.\cr
 & &\left.  \right. \cr
 & &\left.+\frac{1}{2} \omega_{{\cal Y}\mu} \left[(m^2 + \xi_{\cal Y} R)g^{\mu\nu} + \zeta_{\cal Y} R^{\mu\nu} \right]  \omega_{{\cal Y}\nu} \right\} .
\label{vortexaction}
\eea

According to equation (\ref{variationsvortex}), $\phi^i$, $v^m_i$ and $\eta$ transform inhomogeneously under the broken translation, Lorentz transformation and ${\cal Y}$ local variations.  Hence the fields can be transformed to the unitary gauge defined by  $\phi^i = v^m_i =\eta =0$. So doing, the covariant one-forms take a simplified form
\bea
\omega^m &=& dx^\mu e_\mu^{~m} =dx^\mu (\delta_\mu^{~m} + E_\mu^{~m}) \cr
\omega_Z^i &=& dx^\mu  A_\mu^i \cr
\omega_{Ki}^m &=& dx^\mu B_{i\mu}^{~m}\cr
\omega_M^{mn} &=& dx^\mu \gamma_\mu^{mn} \cr
\omega_{\cal Y} &=& dx^\mu B_{{\cal Y}\mu}\cr
\omega_Q &=& dx^\mu a_\mu .
\label{VortexMCOne-formUnitary}
\eea
As before the vierbein reads $e_\mu^{~m} = \delta_\mu^{~m} + E_\mu^{~m}$ and thus depends only on the gravitational fluctuation field, $E_\mu^{~m}$, about the flat background vierbein $\delta_\mu^{~m}$.  

In the unitary gauge the action, equation (\ref{vortexaction}), reduces to that of a massive Proca field $A^i_\mu$ with $Q$ charge $\pm \sin{\hat\theta}$ eigenstates, $A^\pm_\mu = \frac{1}{\sqrt{2}}(A^1_\mu \mp iA^2_\mu)$, and an additional $Q$ neutral massive Proca field $B_{{\cal Y}\mu}$ and finally a massless $U_Q(1)$ gauge field $a_\mu$, all coupled to a gravitational field along with a possible cosmological constant
term so that
\bea
\Gamma &=& \int d^4 x \det{e} \left\{ \Lambda + \frac{1}{2\kappa^2} R -\frac{Z_{\cal Y}}{4}F_{{\cal Y}\mu\nu} F^{\mu\nu}_{{\cal Y}}-\frac{Z_Q}{4}F_{Q\mu\nu} F^{\mu\nu}_{Q} \right. \cr
 & &\left.  \right. \cr
 & & \left. -\frac{Z_{{\cal Y}Q}}{2}F_{{\cal Y}\mu\nu} F^{\mu\nu}_{Q} - \frac{Z_1}{4}F^i_{\mu\nu} F^{i\mu\nu}\right. \cr
 & &\left.  \right. \cr
 & &\left.+\frac{1}{2} A_{\mu}^i \left[(M^2 + \xi R)g^{\mu\nu}\delta_{ij} + \zeta R^{\mu\nu}\delta{ij} + \chi F_{{\cal Y}}^{\mu\nu}\epsilon_{ij} + \rho F_Q^{\mu\nu}\epsilon_{ij} \right]  A_{\nu}^j \right.\cr
 & &\left.  \right. \cr
 & &\left.+\frac{1}{2} B_{{\cal Y}\mu} \left[(m^2 + \xi_{\cal Y} R)g^{\mu\nu} + \zeta_{\cal Y} R^{\mu\nu} \right]  B_{{\cal Y}\nu} \right\} . \label{gravvoract}
\eea
This is the action that describes a gravitating vortex in D=6 dimensional space-time.

\newsection{Discussion}
Standard Model fields can be coupled to the action (\ref{gravvoract}) in an invariant way. In order to specify how the Standard Model fields transform under the full local $ISO(1,5)\times U_{\cal Q}(1) \times SU_C(3) \times SU_W(2) \times U_Y(1)$ symmetry group, each field has to be assigned a charge under the $U_Q(1)$ symmetry apart from its customary Lorentz group representation and Standard Model quantum number assignments. The resulting invariant action describes the interaction of Standard Model fields with the gravitational gauge fields required to realize the full local symmetry group, including the graviton, a massless $U_Q(1)$ gauge field, a pair of massive vector fields that carry $U_Q(1)$ charge, and one massive vector field that is neutral under $U_Q(1)$. Such a low energy effective action and its generalizations as described in this paper allow a systematic study of the phenomenology of braneworld scenarios in a model independent way. The structure of said action depends only on the symmetry breaking pattern and is the universal part of the full low energy effective theory, since it contains those degrees of freedom that are required by symmetry, independent of the details of the underlying short distance physics. The only assumptions are that the action is invariant under a higher dimensional local space-time symmetry group and that the long distance physics can be described by a four dimensional effective field theory, as is consistent with observations.

If the $U_Q(1)$ symmetry exists in addition to the conventional hypercharge symmetry it is interpreted as a $U(1)'$ \cite{Leike:1998wr} with corresponding $Z'$ phenomenology.  Moreover, if the action contains massive fermions beyond those of the Standard Model that are charged under $U_Q(1)$ and at the same time carry conventional hypercharge, then stable particles with a millicharge \cite{Holdom:1985ag} can result. In another, more minimal,  scenario the unbroken $U_Q(1)$ symmetry simply plays the role of the conventional Standard Model hypercharge symmetry. In this case the full local symmetry group is $ISO(1,5)\times U_{{\cal Q}}(1) \times SU_C(3) \times SU_W(2)$. 

The longitudinal components of the massive vector fields $A^{\pm}_\mu$  are the residual consequences of  the Nambu-Goldstone modes that exist in the corresponding globally invariant action due to the spontaneous breaking of two translation symmetries.  It is interesting to note that the occurance of a massive resonance is described in \cite{Shaposhnikov:2005hc} as the remnant of the Nambu-Goldstone boson associated with the sole broken translation symmetry in a specific model with a gravitating domain wall embedded in D=5 dimensions. The presence of  the massive vector fields  $A^{\pm}_\mu$ in the invariant low energy effective action constructed here reflects the necessary occurance of a pair of similar massive resonances in any gravitating vortex model embedded in D=6 dimensions.

Beyond the Abelian Higgs model, vortices also occur in non-Abelian Higgs models. 
These vortices feature so-called non-Abelian orientational moduli 
which are associated with spontaneously broken color-flavor locking symmetry \cite{Hanany:2003hp}.  As a consequence, additional Nambu-Goldstone degrees of freedom are present in the low energy effective world volume theory describing such vortices, as required by the symmetry breaking pattern. 

If defects exist in models with global supersymmetry, then Goldstino fermions associated with spontaneously broken supersymmetry generators are a necessary ingredient of the low energy world volume effective action.
For BPS saturated defects, only a fraction of the supersymmetry is spontaneously broken, while its complement
is realized linearly in the effective theory. Supersymmetric extensions of non-Abelian Higgs models in D=5 and 6 dimensions allow BPS domain walls and vortices, respectively. The effective world volume action for the massless degrees of freedom localized on these defects was considered in the case of global supersymmetry and in the absence of gravity for BPS domain walls embedded in  D=5, N=1 superspace \cite{Isozumi:2004jc} and BPS vortices embedded in D=6, N=1 superspace \cite{Eto:2004ii}. An extension of these actions to the gravitating case is possible by a supersymmetric generalization of the results presented here.

Finally, returning to the general results of section \ref{section3}, the $SO(N)$ gauge fields $B_\mu^{ij}$ can potentially exhibit confinement. The $SO(N)$ representations of the world volume matter content determine the phase of the $SO(N)$ gauge theory. The effective action considered here is valid below the inverse width scale of the soliton.  The mass scale $M$ of the $A^i_\mu$ vector fields associated with the spontaneously broken local translation symmetries is assumed to be significantly below the inverse width, while potentially the confinement scale of the $SO(N)$ gauge theory (or its breaking scale if it is in a Higgs phase) is yet lower again. Hence in case the theory confines, the effective action provides a description of dynamics between the inverse width scale and the confinement scale.


\newpage
\noindent The work of TEC, STL and CX was supported in part by the U.S. Department of Energy under grant DE-FG02-91ER40681 (Task B).  The work of TtV was supported in part by a Cottrell Award from the Research Corporation.  TtV would like to thank the theoretical physics group at Purdue University for their hospitality during his sabbatical leave from Macalester College.

\end{document}